\documentclass[12pt]{article}
\usepackage{epsf}
\usepackage{amsfonts}
\usepackage{subfigure}

\def\COBE{{\rm COBE}}
\def\df{{\rm d}}
\def\e{{\rm e}}
\def\GeV{{\rm GeV}}
\def\H{{\rm H}}
\def\inf{{\rm inf}}

\newcommand{\be}{\begin{equation}}
\newcommand{\ee}{\end{equation}}
\newcommand{\bea}{\begin{eqnarray}}
\newcommand{\eea}{\end{eqnarray}}

\newcommand{\complex}{{{\rm I} \kern -.59em {\rm C}}}

\begin{document}
\begin{titlepage}
  \renewcommand{\thefootnote}{\fnsymbol{footnote}}
  \begin{flushright}
    \begin{tabular}{l@{}}
      IFUNAM-FT00-02\\
      hep-th/0006200
    \end{tabular}
  \end{flushright}
 
  \vskip 0pt plus 0.4fill

  \begin{center}
    \textbf{\LARGE Low scale supergravity inflation with R-symmetry}
  \end{center}

  \begin{center}
    {\large
    G.~Germ{\'a}n$^a$%
    \footnote{E-mail: gabriel@ce.fis.unam.mx},
    A. de la Macorra$^b$%
    \footnote{E-mail: macorra@ft.ifisicacu.unam.mx},
    M.~Mondrag{\'o}n$^b$%
    \footnote{E-mail: myriam@ft.ifisicacu.unam.mx}\\[0.3cm]
    }
    \textit{
  $^{(a)}$Centro de Ciencias F\'{\i}sicas, Universidad Nacional Aut{\'o}noma de M{\'e}xico,\\
  Apartado Postal 48-3, 62251 Cuernavaca, Morelos,  M{\'e}xico\\[0.3cm]
  $^{(b)}$Instituto de F\'{\i}sica, Universidad Nacional Aut{\'o}noma de
  M{\'e}xico,\\
  Apdo. Postal 20-364, 01000 M{\'e}xico D.F. M{\'e}xico}\\
  
    \vspace{1ex}
    \vskip 1ex plus 0.3fill


    \vskip 1ex plus 0.7fill

    \textbf{Abstract}
  \end{center}
  \begin{quotation}
 We study a supergravity model of inflation with R-symmetry and a single scalar
field, the inflaton, slowly rolling away from the origin. The scales of 
inflation can be as low as the supersymmetry breaking scale of $10^{10}$ GeV 
or even the electroweak scale of $10^3$ GeV which could be relevant in the 
context of theories with submillimiter dimensions. Exact analytical solutions 
are presented and a comparison with related models is given.
  \end{quotation}

  \vskip 0pt plus 2fill

  \setcounter{footnote}{0}

\end{titlepage}
 
\section{Introduction}\label{intro}
Recently a model has been studied \cite {ggs} where new inflation \cite {new}
is driven by a slow-rolling inflaton field, characterised by a quadratic 
potential, and incorporating radiative corrections within the context of 
supergravity. The so called $\eta$-problem is dealt with by radiative 
corrections to the inflaton mass $m_{\phi}^2$ which reduce its value from 
the Planck scale. A light inflaton field is confined at the origin by thermal 
effects naturally generating the initial conditions for a (last) stage of new
inflation. Low powers of the inflaton dominate the potential during the era of 
observable inflation thus generating `quadratic' inflation. The nice features of 
this model are that inflation can occur at the scale of supersymmetry breaking 
thus without having to invoke a new scale for inflation. Also the possibility 
of having electroweak scale inflation is realized without any extra difficulty.
To implement this model a superpotential of the hybrid type containing two 
fields was used \cite {ggs}. 

Here, we would like to explore the possibility of obtaining similar results in a
more economical model with a single scalar field. For this an $R$-invariant 
superpotential is proposed in such a way that we can maintain the most important 
conclusions discussed previously in \cite {ggs}.
In particular the $R$-symmetry of the superpotential restricts the powers that 
the inflaton can have. This forbids certain models which occur in 
\cite {ggs} but still maintaining others with low scales of inflation.
 
Analytical solutions can be worked out and a full description of the various 
quantities of interest during the inflationary era is given. In particular we 
find models which allow scales as low as the supersymmetry breaking scale of 
$10^{10}$ GeV or even the electroweak scale of $10^3$ GeV which could be 
relevant in the context of theories with submillimiter dimensions \cite {dimo}.
We also find that the reheat temperature is not sufficiently high in general
thus some other more efficient mechanism should be at work to attain higher 
reheat temperatures.

\section{A Model for Low Scale Inflation} \label{model}

The model we propose to study is given by the following superpotential 
\begin{equation}
 W(\phi)  = \Delta^2 \phi(1-\frac{\kappa}{p+1}\frac{\phi^p}{\Delta^q}),
\label{sp}
\end{equation}
and the K\"{a}hler potential
\begin{equation}
 K(\phi,\phi^*)= \phi\phi^*+\frac{\mu}{4}(\phi\phi^*)^2+...,
\label{kp}
\end{equation}
where p, q are integer numbers and $\mu$ is a constant parameter with
a value fixed by the inflationary constraints. 
The quantities $\kappa$, $\mu$, and $\Delta$ have dimensions of
$M^{q-p}$, $M^{-2}$, and ${M}$ respectively.
From now on we will take $M \equiv M_{\rm
  P}/\sqrt{8\pi} = 1$.  The inflaton superfield
$\phi$ and $\Delta$ have R-charges given by
\begin{equation}
 R\phi(\theta) = \frac{2}{n}, \qquad R\Delta^2 = 2-\frac{2}{n}.
\label{rcharges}
\end{equation}
That is the superfield $\phi(\theta)$ transforms
\begin{equation}
 \phi(\theta)\to \phi^{'}(\theta^{'})  = \e^{i\frac{2}{n}\alpha}\phi
 (e^{-i\alpha}\theta),
\label{fi}
\end{equation}
where $n$ is a positive integer.

The form of Eqs.(\ref{sp})-(\ref{fi}) has been studied before by Izawa
and Yanagida \cite {japon} where they consider a natural inflationary
model in broken supergravity based on an R-symmetry. The new
ingredient in our superpotential is the appearance of the scale of
inflation $\Delta$ in the higher dimension non-renormalizable terms. As
has been discussed at lenght in \cite {ggs}, \cite {progress} these
higher order terms might arise as a result of integrating out heavy
fields in the theory thus generating a mass scale $M^{'}$ in the
denominator much less than the Planck scale. The scale $M^{'}$ can be
associated with any of the scales in the theory in particular with the
inflationary scale simply by writing $M'^{p}=\Delta^{q}M^{p-q}$ (see
Section \ref {compa}). This avoids the introduction of yet another
scale in the model and allows the interesting possibility of
identifying the scale of inflation $\Delta$ with that of supersymmetry
breaking or even with the electroweak scale, of interest for theories
with large extra dimensions \cite {dimo}. As has been shown before
\cite {ggs} the factor $1/\Delta^{q}$ in the higher order terms allows, in
quadratic inflation, practically any scale of inflation.

  The scale $\Delta$ can be though as
due to the presence of a composite superfield which condenses when the
(gaugino) interaction becomes strong at the scale $\Delta$, breaking
the $U(1)_R$ or $Z_n$ symmetry of the model. This R-symmetry specifies
the superpotential and imposes the following relation between p and q
\begin{equation}
 p= \frac{q}{2}(n-1).
\label{p}
\end{equation}
It is the presence of the scale $\Delta$ through the factor
$\Delta^{-q}$ in Eq.(\ref{sp}) which allows to have low scale
inflation as shown below. Also, the $\mu$-parameter appearing in the
K\"{a}hler potential Eq.(\ref{kp}) enters in the mass term for the
inflaton $m_{\phi}^2 \sim \mu \Delta^4$. No other contributions to
$m_{\phi}^2$ occur in the tree level potential \cite {japon0}. To show
this let us consider the supergravity potential \cite {bailin}
\begin{equation}
 V = \exp\left(K\right)
     \left[F^{A\dagger}(K_A^B)^{-1}F_{B} -
     3 |W|^2\right] + {\rm D-terms} ,
\label{pot}
\end{equation}
where
\begin{equation}
 F_{A} \equiv \frac{\partial W}{\partial \Phi^A} +
       \left(\frac{\partial K}{\partial\Phi^A}\right) W ,\qquad
 \left(K_A^B\right)^{-1} \equiv
 \left(\frac{\partial^2
K}{\partial\Phi^A\partial\Phi_B^\dagger}\right)^{-1}.
\end{equation}
For small field values we can expand $V$ so that
\begin{equation}
V \approx \Delta^4(1-\mu \phi^2+\mu^{'} \phi^4-2\kappa\frac{\phi^p}
 {\Delta^q}+\kappa^2\frac{\phi^{2p}}{\Delta^{2q}}+...),
\label{poten}
\end{equation}
where $\mu^{'} = 2-\frac{7}{4}\mu+\mu^2.$ Since $\Delta \ll 1$ the
$\phi^4$ term is much less than $\phi^p/\Delta^q$ for $p=4$. For $p >
4$, $\phi^4 \ll \phi^p/\Delta^q$ whenever $\phi \gg
\Delta^{\frac{q}{p-4}}$ which is always the case in the examples of
interest we study below.  When $\phi$ is much less than one, higher
order terms in $\phi$ are negligible, and have been omitted in
Eq.(\ref{poten}). In this case we can work with the simpler expression
\begin{equation}
 \frac{V}{\Delta^4} = \left(1-\kappa \frac{\phi^p}{\Delta^q}\right)^2
 -\mu \phi^2, 
\label{potential}
\end{equation}
which is practically indistinguishable from the full supergravity potential
Eq.(\ref{pot}) all the way to the global minimum.

\section{Analytical Solutions}\label{ana}

Here we obtain closed form expressions for the relevant quantities involved
in the inflationary era. We are assuming that the radiative corrections to
the inflaton mass $\sim \ln \phi$ are already included in the parameter $\mu$
and we take $\mu$ fixed by its value at $\phi_{\H}$ (where the subscript $\H$ 
denotes the epoch at which a fluctuation of wavenumber $k$ crosses the Hubble 
radius $H^{-1}$ during inflation). This is not a great sin since the $\ln\phi$
corrections change very slowly from $\phi_{\H}$ to the end of inflation at 
$\phi_e$ and it turns out to be a very good approximation \cite {progress} to 
consider $\mu$ as a constant. The advantage of doing this is that we can obtain 
\cite {progress} closed form solutions. The parameter $\mu$ can take positive or 
negative values. In particular when $\mu<0$ there is a maximum at
\begin{equation}
\phi _{max}\approx \left(\frac{-\mu \Delta ^q}{\kappa p} \right)^{\frac{1}{p-2}},
\label{fimax}
\end{equation}
when $\mu \to 0, \phi_{max} \to 0$ as it should. In this case
the $-\mu \phi^2$-term dominates $V(\phi)$ in the interval 
$0\leq \phi \leq\phi_{max}$. Inflation for $\phi>\phi_{max}$ requires the
participation of both $-\mu \phi^2$ and $-2\kappa \phi^{p}/\Delta^{q}$ with 
the last term dominating during inflation. Thus we cannot talk about
"quadratic" inflation when $\mu<0$, this can only occur for positive $\mu$. The 
following expressions, however, are valid for any $\mu$.

1) {\bf The end of inflation.} In the models under consideration
inflation is generated while $\phi $ rolls to larger values. The end of
inflation occurs at $\phi =\phi _{e}$ when the slow roll conditions 
\cite{review} are
violated. The slow-roll conditions are upper
limits on the normalised slope and curvature of the potential:
\begin{equation}
 \epsilon \equiv \frac{1}{2}\left(\frac{V'}{V}\right)^2 \ll \gamma ,\qquad
 |\eta| \equiv \left|\frac{V''}{V}\right| \ll \gamma .
\label{slowroll}
\end{equation}
The potential determines the Hubble parameter during inflation as
$H_\inf\equiv\dot{a}/a\simeq\sqrt{V/3M^2}$.  Inflation ends
(i.e. $\ddot{a}$, the acceleration of the cosmological scale factor,
changes sign from positive to negative) when $\epsilon$ and/or
$|\eta|$ become of ${\cal O}(\gamma)$. This occurs at 
$V^{\prime \prime }(\phi )\approx -\gamma $, where
$\gamma ={\cal O}(1).$ Thus we have 
\begin{equation}
\phi _{e}\approx \left( \frac{(\gamma -2\mu)\Delta ^{q}}{2\kappa p(p-1)}%
\right) ^{\frac{1}{p-2}}.
\label{fie}
\end{equation}

2) {\bf Scalar density perturbations.} The adiabatic scalar density 
perturbation generated through quantum
fluctuations of the inflaton is \cite{review}
\begin{equation}
 \delta^2_\H (k) = \frac{1}{150\pi^2} \frac{V_\H}{\epsilon_\H}\ ,
\label{deltah}
\end{equation}
where the subscript $\H$ denotes the epoch at which a fluctuation of
wavenumber $k$ crosses the Hubble radius $H^{-1}$ during inflation,
i.e. when $aH=k$. (We normalise $a=1$ at the present epoch, when the
Hubble expansion rate is $H_0\equiv100h$~km\,s$^{-1}$Mpc$^{-1}$, with
$h\sim0.5-0.8$). The COBE observations \cite{cobe} of anisotropy in
the cosmic microwave background on large angular-scales require
\cite{review}
\begin{equation}
 \delta_\COBE \simeq 1.9 \times 10^{-5} ,
\label{cobe}
\end{equation}
on the scale of the observable universe
($k_\COBE^{-1}\sim\,H_0^{-1}\sim3000h^{-1}$~Mpc). In addition, the
COBE data fix the spectral index,
$n_\H(k)\equiv1+\df\delta_\H^2(k)/\df\ln\,k=1-6\epsilon_\H+2\eta_\H$,
on this scale:
\begin{equation}
 n_\COBE = 1.2 \pm 0.3 .  \label{ncobe}
\end{equation}
Solving Eq. (\ref{deltah}) we find 
\begin{equation}
\phi _{\H}^{p-1}+\frac{\mu\Delta ^{q}}{\kappa p}\phi _{\H}-\frac{\Delta
^{q+2}}{2\kappa pA_{\H}}=0,
\label{efih}
\end{equation}
where $A_{\H}\equiv \sqrt{75}\pi \delta _{\H}.$ This equation determines $\Delta $
once $\phi _{\H}$ is determined.

3) {\bf Number of e-folds.} The number of e-folds from $\phi_{\H}$ to the
end of inflation at $\phi_e$ is
\bea
 N_{\H} &\equiv& -\int_{{\phi}_{\H}}^{{\phi}_\e}
         \frac{V({\phi})}{V'({\phi})} \df{\phi}
\approx \int\limits_{\phi _{H}}^{\phi _{e}}d\phi \frac{1}{2\mu\phi +2\kappa
p\phi ^{p-1}/\Delta ^{q}} \nonumber\\
&=&-\frac{1}{2\mu(p-2)}\ln \left( \frac{1+\frac{\mu\Delta
^{q}}{\kappa p\phi _{e}^{p-2}}}{1+\frac{\mu\Delta ^{q}}{\kappa p\phi
_{\H}^{p-2}}}\right).
\label{ene}
\eea
Solving for $\phi _{\H}$ gives 
\begin{equation}
\phi _{\H}=\left( \frac{-\mu\Delta ^{q}}{\kappa p(1-(1+\frac{2\mu(p-1)}{\gamma
-2\mu%
})e^{2\mu(p-2)N_{\H}})}\right) ^{\frac{1}{p-2}}\equiv B\Delta ^{\frac{q}{p-2}}.
\label{fih}
\end{equation}
Finally substituting in Eq.(\ref{efih}) and simplifying we obtain the
required solution for $\Delta $ 
\begin{equation}
\Delta =\left( 2\kappa pA_{\H}(B^{p-1}+\frac{\mu B}{\kappa p})\right) ^{\frac{%
p-2}{2(p-2)-q}}.
\label{del}
\end{equation}

4) {\bf Spectral Index}. We now readily obtain a form for the spectral
index 
\begin{eqnarray}
n_{\H} &\approx &1+2V^{\prime \prime }(\phi _{\H})  \nonumber \\
&\approx &1-4\mu-4\kappa p(p-1)B^{p-2}.
\label{si}
\end{eqnarray}

5) {\bf Reheat temperature.} One obvious effect of lowering the scale of the 
inflationary potential is the lowering of the reheat temperature. At the end of
inflation the oscillations of the inflaton field would make it decay thus 
reheating the universe.
The couplings of the inflaton to  some other bosonic $\chi$ or
fermionic $\psi$ MSSM fields occur due to terms
$-\frac{1}{2}g^2\phi^2\chi^2$ or $-h\bar{\psi}\psi\phi$, respectively. These
couplings induce decay rates of the form \cite{inflbook}
\begin{equation}
\Gamma(\phi\to\chi\chi)=\frac{g^4\phi_0^2}{8\pi m_{\phi}},\qquad
\Gamma(\phi\to\bar{\psi}\psi)=\frac{h^2m_{\phi}}{8\pi}
\label{decays}
\end{equation}
where $\phi_0$ is the value of $\phi$ at the minimum of the potential
\begin{equation}
\phi_0 \approx \left( \frac{\Delta^q}{\kappa}\right)^{\frac{1}{p}},
\label{fi0}
\end{equation}
and $m_{\phi}$ is the inflaton mass given by
\begin{equation}
m_{\phi}\approx\sqrt {2} p \kappa^{\frac{1}{p}}\Delta^{2-\frac{q}{p}}.
\label{inflamass}
\end{equation}
A maximum value for the decay is obtained when $m_{\chi,\psi}\approx
m_{\phi}$. In this case we find
\begin{equation}
\Gamma\approx \frac{m_{\phi}^3}{8\pi\phi_0^2}
\label{decay}
\end{equation}
The reheat temperature at the beginning of the radiation-dominated era is thus
\cite{turner}
\begin{equation}
T_{rh} \approx \left( \frac{90}{\pi^2g_{*}}\right)^{\frac{1}{4}} 
min\left(\sqrt{H(\phi_e)},\sqrt{\Gamma}\right)
\approx \left(\frac{30}{\pi^2g_{*}}\right)^{\frac{1}{4}}
min\left(\Delta,\left( \frac{3}{8\pi^2}\right)^{\frac{1}{4}}p^{\frac{3}{2}}
\kappa^{\frac{5}{2p}}\Delta^{3-\frac{5q}{2p}}\right)
\label{trh}
\end{equation}
where $g_*$ is the number of relativistic degrees of freedom which for
the MSSM equal $915/4$.

6) {\bf Quantum fluctuations.} The value $\phi_{\H}$ at the beginning of 
the last $N_{\H}$ e-folds of inflation should exceed the quantum fluctuations 
of the inflaton $\delta\phi \approx \frac{H}{2\pi} \approx \frac{\Delta^2}
{2\pi\sqrt{3}}$. From Eqs.(\ref{fih}) and (\ref{del}) we can impose the 
following condition
\begin{equation}
\frac{\delta\phi}{\phi_{\H}}\approx  \frac{\Delta^{\frac{2(p-2)-q}{p-2}}}
{2\pi\sqrt{3} B} \approx  (\mu+\kappa p B^{p-2})\times10^{-4}\ll 1.
\label{qf}
\end{equation}
Typically $B\leq 10^{-2}$ and the condition Eq.(\ref{qf}) is easily verified 
by the models we are interested in.

\section{Numerical Results}\label{numerical}

The number $N_{\H}$ of $e$-folds of the present horizon is given by 
\cite {number}
\begin{equation}
N_{\H}\approx 67+\frac{1}{3}\ln H+\frac{1}{3}\ln T_{rh}
\approx
67+\frac{1}{3}\ln\left(\left(\frac{10}{3\pi^2g_{*}}\right)^{\frac{1}{4}}
min\left(\Delta^3,\left( \frac{3}{8\pi^2}\right)^{\frac{1}{4}}p^{\frac{3}{2}}
\kappa^{\frac{5}{2p}}\Delta^{5-\frac{5q}{2p}}\right)\right).
\label{efolds}
\end{equation}
Solving this equation consistently with Eq.(\ref{del})
we can obtain a representative set of values for the various
quantities of interest during inflation. A sample is given in Table 1.
\begin{table}[tbp] 
%
\begin{tabular}{|c|c|c|c|c|c|c|c|c|c|}
\hline
{\bf p} & {\bf q} & ${\bf \kappa}$ & ${\bf \mu}$ & ${\bf \phi_H}$ & ${\bf
  \phi_e}$ & ${\bf n_H}$ & ${\bf N_H}$ & ${\bf \Delta  (GeV)}$ & ${\bf
  T_{rh}  (GeV)}$  \\  
\hline
4 & 2 & 1 & 0.024 & 1.4 10$^{-9}$ & 3.7 10$^{-8}$ & 0.90 & 48 & 4.6
10$^{11}$ & 4.9 10$^{6}$ \\  
5 & 5 & 1 & 0.024 & 1.1 10$^{-31}$ & 7.0 10$^{-31}$ & 0.90 & 25  & 4.2 &
1.4  \\  
5 & 5 & 1.8 10$^{-3}$ & 0.024 & 6.7 10$^{-27}$ & 5.4 10$^{-26}$ & 0.90 & 31 &
10$^{3}$ & 340   \\  
8 & 8 & 2.4 10$^{-3}$ & 0.024 & 6.7 10$^{-13}$ & 8.1 10$^{-12}$ & 0.90 & 47 &
10$^{10}$ & 3.4 10$^{9}$  \\
\hline 
\end{tabular}
\caption{A sample set of values for quadratic inflation for the model
  defined by the superpotential $W(\phi) = \Delta^2
  \phi(1-\frac{\kappa}{p+1}\frac{\phi^p}{\Delta^q})$  
and the K{{\"a}}hler potential $K(\phi,\phi^*)=
\phi\phi^*+\frac{\mu}{4}(\phi\phi^*)^2+...$. 
The quantities $\kappa,~\mu$ (with dimensions of $M^{q-p}$ and $M^{-2}$
respectively), and $\phi$  are normalised by the Planck mass.
\label{table:1}}%
\end{table}%
%

\begin{figure}[t]
\centering
\mbox{\subfigure[]{\epsfxsize=10cm\epsfysize=10cm
\epsffile{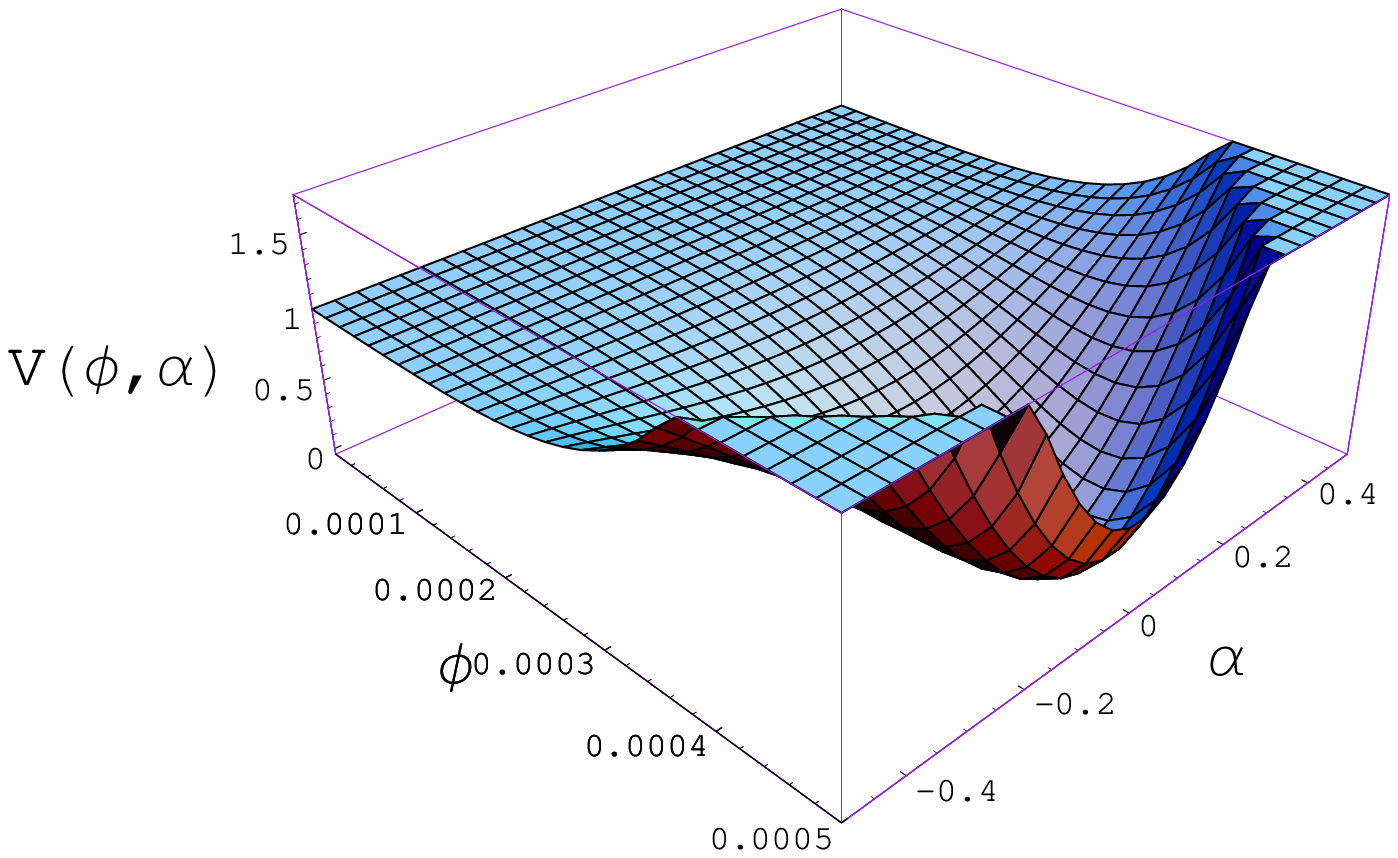}}
  
\subfigure[]{\epsfxsize=6cm\epsfysize=6cm
\epsffile{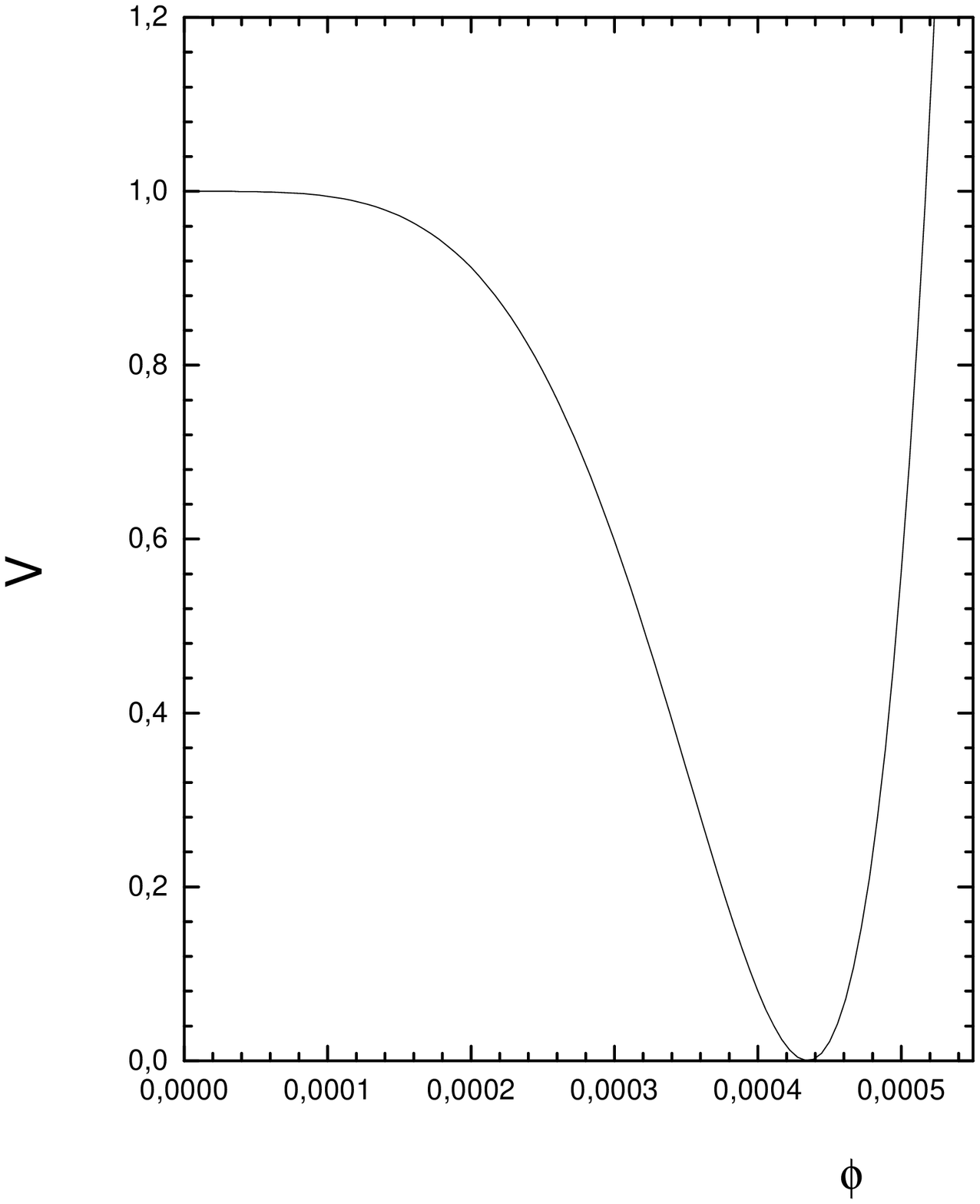}}}
\caption{(a) The inflationary potential (in units of $V_0\equiv\Delta^4$) 
  Eq.(\ref{pot}),
  is shown as a function of the magnitud of $\phi$, denoted again by
  $\phi$, and
  its phase $\alpha$ for the case $(p,q)=(4,2)$,
  $\mu=0.024$ and ${\kappa}=1$. We notice that $\alpha=0$ is a stable
  direction of the potential. The height of the potential
  corresponds to a scale of $4.6 {\times} 10^{11}$ GeV.\newline
 (b) Notice that
  $<\phi_0 >\sim 10^{-4} \ll 1$, thus Eq.(\ref{potential}) remains a
  good approximation for the whole potential.}
\label{fig1}
\end{figure}

\begin{figure}
\epsfxsize=13cm\epsfysize=12cm
\centerline{\epsffile{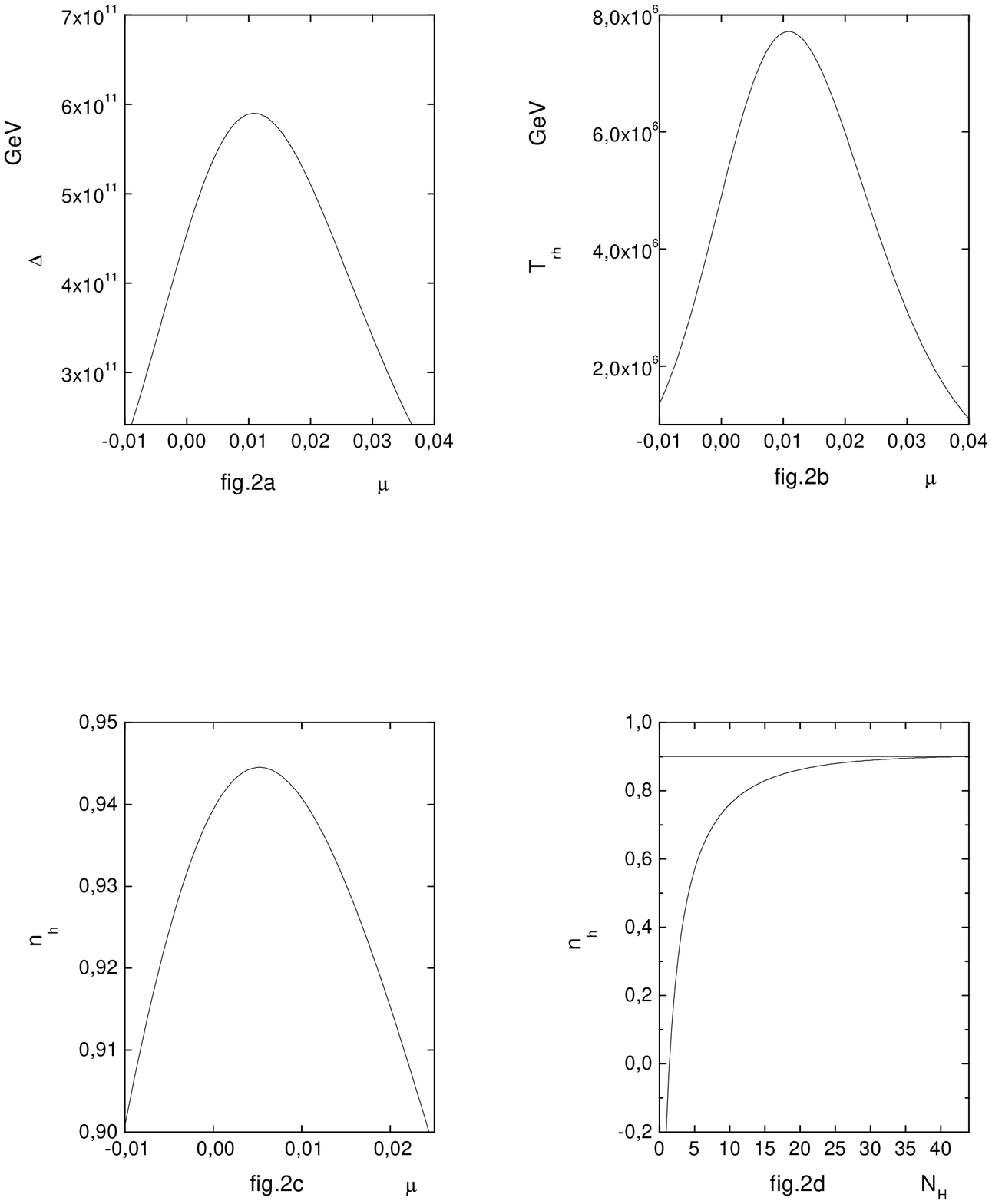}}
\caption{The scale of inflation $\Delta$ given by Eq.(\ref{del})
  is shown in Fig. 2a as a function of $\mu$ for the case
  $(p,q)=(4,2)$, $N_{\H}=48$ and $\kappa=1$.  Similar behaviour occurs for
  the other $(p,q)$ cases although with different maximum values.  
  Fig. 2b: The reheat temperature $T_{rh}$ given by Eq.(\ref{trh}) is
  shown as a function of $\mu$ for the case $(p,q)=(4,2)$, $N_{\H}=48$
  and $\kappa=1$.
  Fig. 2c: The spectral index Eq.(\ref{si}) is shown as a function of
  $\mu$ for the case $(p,q)=(4,2)$, $N_{\H}=48$ and $\kappa=1$. Note that
  there is a maximum value which $n_{\H}$ can have. The same behaviour
  occurs for the other $(p,q)$ cases although with different maximum
  values.
  Fig. 2d: The spectral index Eq.(\ref{si}) is shown as a function
  of the number $N_{\H}$ of e-folds of inflation from the end of
  inflation as the origin, where it takes a value $-1$. This figure
  corresponds to the case $(p,q)=(4,2)$, $\mu=0.024$ and $\kappa=1$.The
  horizontal line correspond to $n_{\H}=0.9$.  Similar shapes are
  found in the other $(p,q)$ cases.}
\label{fig2}
\end{figure}

We now plot in $Fig.1$ the inflaton potential $V(\phi,\alpha)$ as a function
of $\phi$ and the phase $\alpha$.  In $Figs. 2a,2b$, and $2c$ we plot the
scale of inflation $\Delta$, the reheating temperature $T_{rh}$, and the
spectral index $n_{\H}$ as functions of the parameter $\mu$,
respectively.  Finally $Fig.2d$ shows the behaviour of the spectral
index as a function of the number $N_{\H}$ of e-folds of inflation
from the end of inflation. All of these figures are for the case
$(p,q)=(4,2)$.  Similar behaviour is found in the other $(p,q)$ cases.

\section{Comparison with related work}\label{compa}

In the models studied in \cite {ggs} and further elaborated in
\cite {progress} quadratic inflation is implemented through a hybrid
mechanism with the participation of two fields.
A linear term in a field $Y$ follows 
if $Y$ carries non-zero $R$-symmetry charge under an unbroken $R$-symmetry.
The inflaton $\phi$ is a singlet under the $R$-symmetry but carries a charge 
under a discrete $Z_{p}$ symmetry. Then the superpotential has the form
\begin{equation}
 W = \left(\Delta^2 - \frac{\phi^p}{M'^{p-2}}
     - \frac{\phi^{2p}}{M'^{2p-2}} - \dots \right)Y.
\end{equation}
This gives rise to the potential
\begin{equation}
 V = \left(\Delta^2 - \frac{\phi^p}{M'^{p-2}}
     - \frac{\phi^{2p}}{M'^{2p-2}} - \dots \right)^2 ,
\label{einf}
\end{equation}
displaying the possibilities of ending inflation.  There are also
terms involving $Y$ which are dropped as they do not contribute to the
vacuum energy since $Y$ does not acquire a vacuum expectation value.
The scale $M'$ denotes new physics below the Planck scale and we can
write $M'^{p-2}=\Delta^{q}M^{p-q-2}$ to take into account  the
possibility that the scale associated with the higher dimension
operators may be below the Planck scale.

In the present model there is only a single scalar field with a
superpotential determined by the $R$-symmetry as shown in
Section~\ref{model}. As a consequence of this symmetry some $(p,q)$
models which occur in \cite{ggs}, \cite {progress} (for example
$(p,q)=(4,3)$) are not allowed here. It is therefore interesting that
most of the results and conclusions of \cite {ggs} are still
maintained.

Other studies of quadratic inflation have concentrated on the case
where radiative corrections make the potential develop a {\em maximum}
near the origin, from which the inflaton rolls either away from the
origin or towards it \cite{quad}, and inflation ends through a hybrid
mechanism.

There is also related work \cite{japon} with a superpotential (in
our notation)
\begin{equation}
 W(\phi)  = \Delta^2 \phi-\frac{\kappa}{n}\phi^{n},
\label{sup}
\end{equation}
and the K\"{a}hler potential
\begin{equation}
 K(\phi,\phi^*)= \phi\phi^*+\frac{\mu}{4}(\phi\phi^*)^2+...,
\label{kap}
\end{equation}
where $\phi$ and $\Delta$ have R-charges as in Eq.(\ref{rcharges}).
However the fact that the factor $\Delta^{-q}$ appearing in our  
Eq.(\ref{sp}) is not present in Eq.(\ref{sup}) above eliminates the
possibility of having low scales for inflation (in \cite{japon} the
lowest scale allowed is ${\cal O}(10^{12}GeV)).$

\section{Conclusions}\label{con}

We have studied a model of inflation where low inflationary scales are
allowed without having to introduce unnatural values for the
parameters involved. The model is defined in terms of a single scalar
field, the inflaton, and the term driving (new) inflation is quadratic
in $\phi$. The end of inflation due to higher order non-renormalisable
terms.  Radiative corrections to the inflaton mass reduce $m_{\phi}^2$
from its natural value at the Planck scale.  For a light inflaton
thermal initial conditions can naturally place the inflaton at the
origin, initiating a (last) stage of new inflation.  A quadratic
parameterisation of the inflationary potential allows low values for
the inflationary scale $\Delta$. One can have
$\Delta\sim10^{10}~\GeV$, the supersymmetry breaking scale in the
hidden sector or the electroweak scale $\Delta\sim10^3~\GeV$ which
could be relevant in the context of theories with submillimeter
dimensions. The well justified assumption that the inflaton mass
parameter $m_{\phi}^2 \sim \mu \Delta ^4$ remains practically constant
during inflation allows analytical closed form expressions for all the
relevant quantities.

\section{Acknowledgements}
G.G. would like to thank G.G. Ross and S. Sarkar for useful
discussions.  This work was supported by the projects PAPIIT IN110200,
and Conacyt 32415-E.

\end{document}